\documentclass[10pt,journal,compsoc]{IEEEtran}
\usepackage{CJK}
\usepackage{amssymb}
\usepackage{amsmath}
\usepackage{graphicx, subfigure}
\usepackage{url, cite}
\usepackage{verbatim}
\usepackage{booktabs}
\usepackage{multirow}
\usepackage{bigstrut}
\usepackage{booktabs}
\usepackage{epstopdf}

\usepackage{amsmath,cite,graphicx,times,subfigure,url,verbatim}
\usepackage{algorithmic}
\usepackage{algorithm}

\usepackage{multirow}

\usepackage{bbm}

\usepackage{booktabs}
\usepackage{float}

\newtheorem{subsec:coding}{subsec:coding}

\usepackage{color}

\begin{document}

\title{Human-Like Hybrid Caching in Software-defined Edge Cloud}

\author{
Yixue Hao, Miao Li, Di Wu, Min Chen, Mohammad Mehedi Hassan, Giancarlo Fortino
\IEEEcompsocitemizethanks{
\IEEEcompsocthanksitem Y. Hao, M. Li and M. Chen are with School of Computer Science and Technology, Huazhong University of Science and Technology, Wuhan 430074, China. (minchen2012@hust.edu.cn)
\IEEEcompsocthanksitem D. Wu is with the Department of Computer Science, School of Data and Computer Science, Sun Yat-sen University, Guangzhou 510006, China. (wudi27@mail.sysu.edu.cn)
\IEEEcompsocthanksitem M. M. Hassan is with College of Computer and Information Sciences, King Saud University, Riyadh 11543, Saudi Arabia, and also with Research Chair of Smart Technologies, King Saud University, Riyadh 11543, Saudi Arabia. (mmhassan@Ksu.edu.sa)
\IEEEcompsocthanksitem  G. Fortino is with University of Calabria, Rende, 87036, Italy. (giancarlo.fortino@unical.it)
}
}

\markboth{Under review: IEEE Internet of Things, VOL. XX, NO. YY, MONTH 20XX}{}

\IEEEtitleabstractindextext{
\begin{abstract}
With the development of Internet of Things (IoT) and communication technology, the number of next-generation IoT devices has increased explosively, and the delay requirement for content requests is becoming progressively  higher. Fortunately, the edge-caching scheme can satisfy users' demands for low latency of content. However, the existing caching schemes are not smart enough. To address these challenges, we propose a human-like hybrid caching architecture based on the software defined edge cloud, which simultaneously considers the content popularity and the fine-grained user characteristics. Then, an optimization problem with a caching hit ratio as an optimization objective is formulated. To solve this problem, using reinforcement learning, we design a human-like hybrid caching algorithm. Extensive experiments show that compared with popular caching schemes, human-like hybrid caching schemes can improve the cache hit ratio by 20\%.
\end{abstract}

}

\maketitle

\IEEEraisesectionheading{\section{Introduction}
\label{sec:introduction}}

\begin{figure*}
\centering
\includegraphics[width=5.6in]{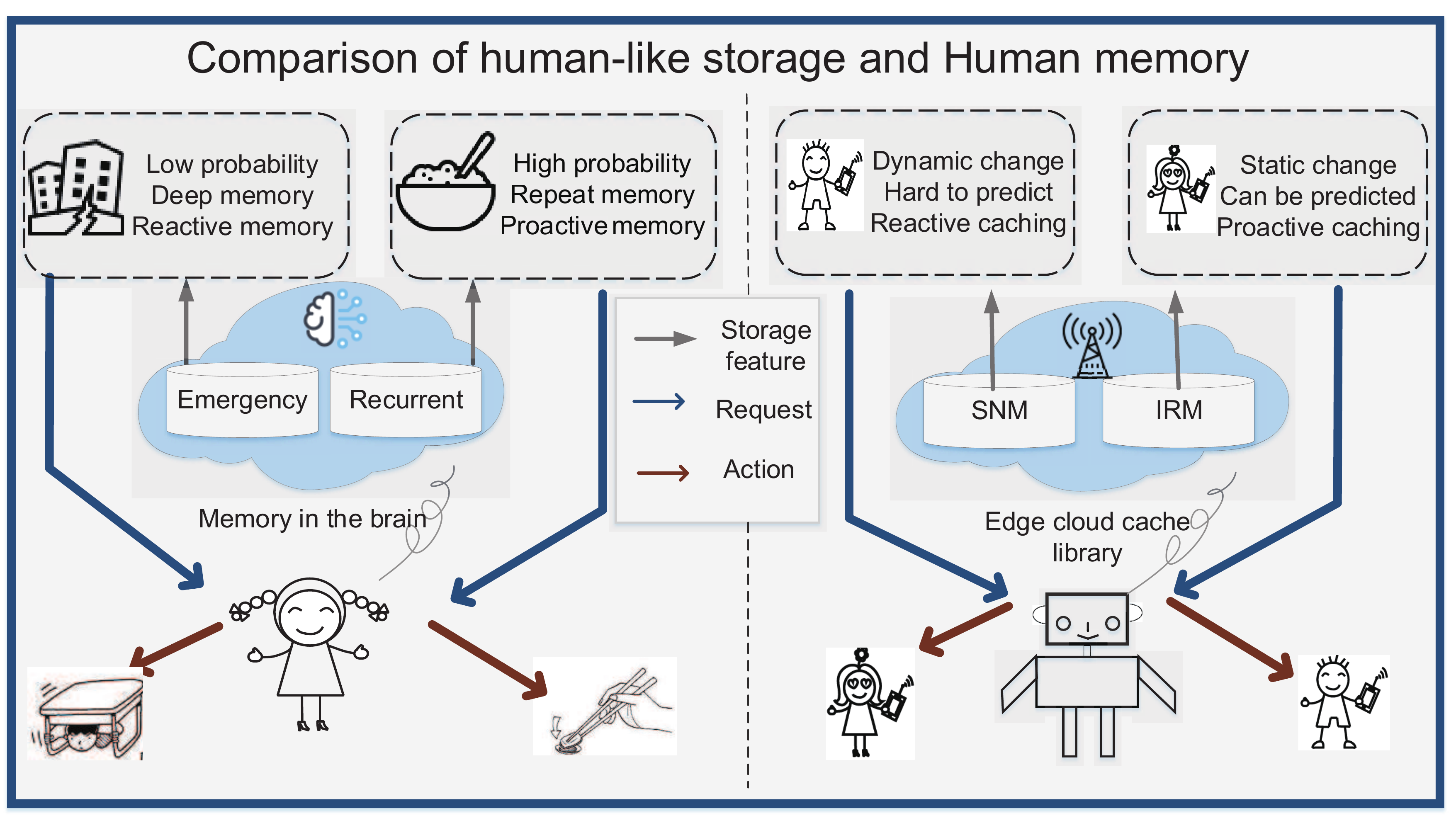}
\caption{Illustration of the concept of human-like hybrid caching architecture in software-defined edge cloud.}
\label{fig01}
\end{figure*}

With the explosive growth of next-generation Internet of Things (IoT) applications, the delay requirements of users for content are becoming increasingly higher, such as virtual reality application~\cite{1,fortino2008using}. However, due to the distance between the content server and users, high communication delay is caused, which can not meet the user's delay requirements~\cite{sun2019tide}. Fortunately, with the introduction of edge cloud, content can be cached on edge cloud to satisfy the user's quality of service (QoE)~\cite{chen2016opportunistic,fortino2014agent}. This is because the edge cloud is very close to the user. Thus, research on edge caching has become one of the most popular research topics in the field of wireless communication~\cite{4,5,6}. Based on the content popularity, popular content can be cached on the edge cloud, which can effectively reduce congestion in the network link and request delay, thereby improving the user's QoE.

However, because the content popularity is dynamically changing, it is difficult to predict the content popularity, which results in the great challenge to find the optimal wireless edge caching strategy. At present, most current research use a single model to analyze the popularity of  content. For example, Hefeeda et al.~\cite{2} show that the popularity of content conforms to an independent reference model (IRM). This model is widely used in most research to study the content popularity. However, the IRM does not take into account the dynamic nature of content popularity.
Thus, not all user content requests conform to the IRM. To better describe content popularity, Ahmed et al.~\cite{3} propose the shot noise model (SNM) by studying the dynamic nature of user requests, and prove the correctness of SNM using real datasets. Some studies also try to design wireless edge caching strategies using the SNM model~\cite{15}, and the cache hit ratio is higher than IRM.

For content popularity prediction, traditional research uses statistical information to make prediction. For example, Tan et al.~\cite{4} use a hypothetical test to study the distribution of user popularity preferences, and propose optimization models for different objectives to  improve the gain of the proactive caching. Moreover, Golrezaei et al.~\cite{5}established a caching optimization problem to minimize the expected download time of files using caching helpers. Guo et al.~\cite{6} design a collaborative local caching strategy based on the heterogeneous file preferences. Christian et al.~\cite{7} propose the use of convolutional neural network to study classification and popularity prediction of music video, and design an proactive caching strategy which reduces the delay. However, these schemes are based on the known static distribution of popularity. In practice, the content popularity can not be known in advance, and the prediction algorithm is not necessarily able to successfully predict the popularity.

In consideration of time-varying characteristics in user content requests, Jiang et al.~\cite{8} propose an online user preference prediction algorithm that takes into account users' temporal users' temporal and spatial contents popularity in the fog radio access network. This algorithm can predict the future contents popularity in an area online (without being restricted by any type of contents) and track the change of popularity in real-time.
Bharath et al.~\cite{9} propose an content popularity estimation method based on transfer learning in heterogenous small cell networks.
Farahat et al.~\cite{12} utilize entropy for mobility management to determine the best proactive cache node and thus eliminate request redundancy, but this scheme leads to higher latency. However, these scheme do not take into account fine-grained characteristic of the content.

Furthermore, the fine-grained characteristic of the content requested by a user also exerts influence on the caching strategy. For example, the size of the content, the bandwidth requirement for the content transmission and the value of content~\cite{chen2019cognitive} are important factors that affect edge cloud caching decisions. Some work has done preliminary research on this. For example, Muller et al.~\cite{10} propose a proactive caching strategy based on context awareness. The algorithm can improve the cache hit ratio by observing the user's context information periodically and updating the cache contents. Spyropoulos et al.~\cite{11} determine the influence of user location on cached data through research on YouTube video data. Although the above research scheme considers the impact of user context information on the design of the caching scheme, it considers that user requests are only single time-varying.

Thus, the content caching strategy described above considers a single content popularity model and does not take into account fine-grained content caching characteristic. The caching strategy is not smart enough.
Therefore, in this paper, we propose a human-like hybrid caching strategy based on the human cognitive memory model (i.e., combines recurrent and emergent cognitive memory with fine-grained characteristic of events).
Specifically, we use the idea of software defined network (SDN) to implement the human-like hybrid content caching. In the software-defined edge cloud, the control plane is similar to the human brain, which can optimize the global content cache based on content fine-grained characteristic and popularity. That is, the controller periodically receives requests from the data plane and makes caching decisions based on learning about the data.

Furthermore, based on the software-defined edge cloud, we first propose a human-like hybrid caching architecture. In this architecture, we consider not only the content popularity (i.e, the user request that conforms to IRM of static distribution and the user request that conforms to SNM of dynamic distribution), but also the influence of the fine-grained characteristics of content. Then, we present an optimization problem aiming at maximizing the cache hit ratio and solve it using human-like hybrid caching algorithm. Finally, experiments are given to verify the human-like hybrid caching scheme.

The main contributions of this paper can be summarized as follows:
\begin{itemize}
\item Based on the software-defined edge cloud, we first propose the concept of human-like hybrid caching. Then, we give the human-like hybrid caching architecture which includes data collection module and human-like learning module. In the learning module, we consider not only the learning of content popularity (i.e., IRM model and the SNM model), but also the learning of fine-grained characteristic of users.

\item We formulate an optimization model to maximize the cache hit ratio based on the human-like hybrid caching architecture. To solve this optimization problem, using multi-arm bandit theory, we propose human-like hybrid caching algorithm.

\item Extensive simulations are carried out to evaluate the performance of the human-like hybrid caching scheme. The results show that the cache hit ratio of the human-like hybrid caching scheme increased by about 40\% and 20\% compared with the random caching algorithm and the popular caching algorithm under different file library size and cache capacity.
\end{itemize}

The rest of this paper is organized as follows: In Section II, we introduce the human-like caching architecture. The system model and problem formulation are described in Section III. In Section IV, we propose a human-like hybrid caching algorithm to solve the optimization problem. The simulation results are shown in Section V. And finally, Section VI concludes the paper and delineates future work.

\section{Human-like caching architecture in Software-defined Edge Cloud} \label{sec:Human-likecaching architecture based on SDN}

In this section, we introduce the human-like hybrid caching architecture based on the software-defined edge cloud.

\subsection{The Concept of Human-Like Hybrid Caching}

The human-like hybrid caching concept is shown in Fig.~\ref{fig01}. Specifically, human-like caching is established by imitating the memory system of the human brain. Human memory is divided into two categories, i.e., emergency memory and recurrent memory ~\cite{aben2012distinction,chen2019label}. Emergency memory is the proactive memory of people for an event in an emergency. The number of these events is small because they have a very low probability of occurrence. However, these events are unforgettable and conform to the characteristics of deep memory. For example, if a person experiences an earthquake, and some tips on earthquake prevention have helped him in this earthquake, then he will passively remember these tips and these tips are unforgettable to him. Moreover, if he experienced another earthquake, he can not help recalling these tips. Recurrent memory refers to events that occur repeatedly. The utilization of chopsticks is an example of this because people often use chopsticks in their daily life, and they remember this event proactively. Events in recurrent memory occur with a high probability, people proactively remember them.

For edge cloud, the popularity of user requested content is taken as an example to describe the human-like hybrid caching based on software-defined edge cloud. Specifically, we divide the cached file library in the edge cloud into IRM library and SNM library. The contents cached in the IRM library correspond to static user requests, which comply with a Zipf distribution and are easy to predict. While the contents that are cached in the SNM library correspond to dynamic user requests are difficult to predict, and they are temporary. The proposed human-like caching system can be viewed as a human-like agent on an edge cloud. When a user request is sent to this to this agent, this agent checks whether the contents being requested is cached on the edge cloud, and the contents will be sent to the user if it is.

\begin{figure}
\centering
\subfigure[The flow of human-like hybrid caching.]
{\includegraphics[width=3.5in]{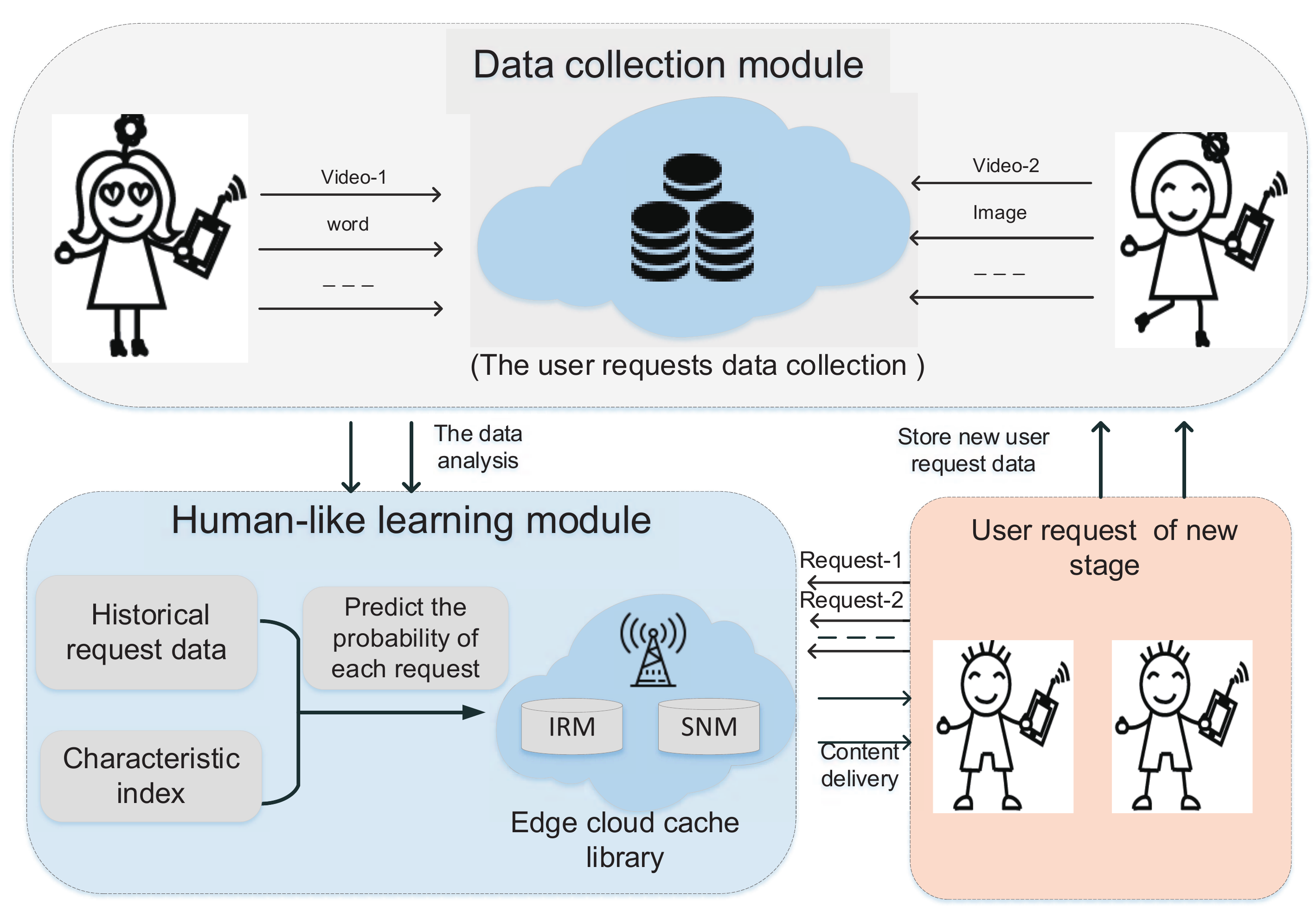}
\label{fig02a}
}
\subfigure[Human-like hybrid caching learning model.]
{\includegraphics[width=3.5in]{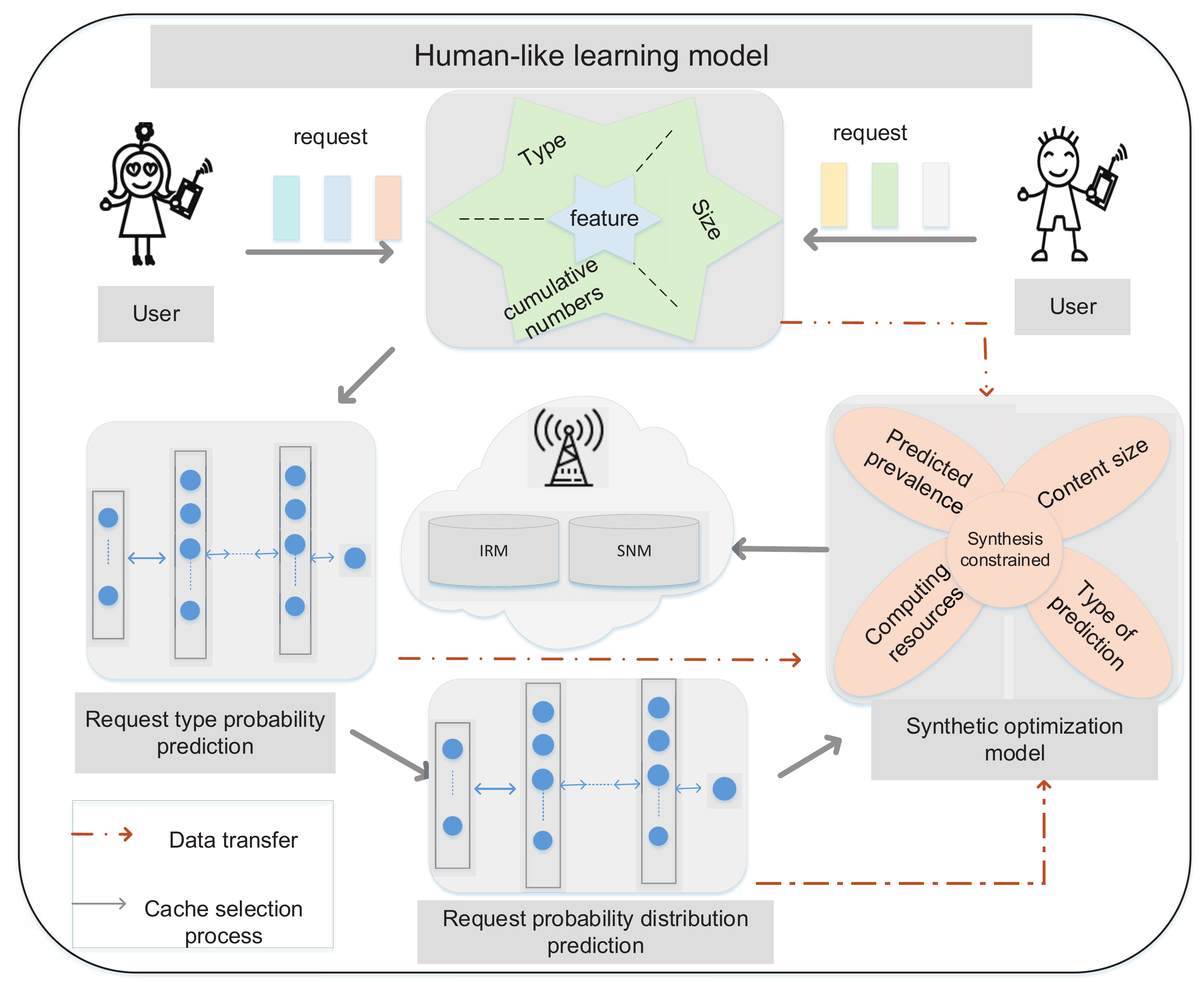}
\label{fig02b}
}
\caption{Human-like learning architecture in software-defined edge cloud.}
\label{fig02}
\end{figure}

\subsection{Human-like Hybrid Caching Architecture }

\subsubsection{The software defined edge cloud}

Nowadays, the SDN technology has been widely used in various fields. Based on the SDN technology, the control plane and the data plane can be separated. In this paper, in order to implement the human-like hybrid caching strategy, we utilize the software defined edge cloud consisting of three parts, i.e., data plane, control plane and user plane.
The data plane is responsible for the collection of users content require and edge cloud resources (including communication, computing and storage resources). The control plane is composed of SDN controller which is used to process the data collected by data plane. The user plane can learn  user characteristics (e.g., user mobility).

Furthermore, in the software-defined edge cloud, the data plane receives the requests for content and the status of network resources, and the controller of the control plane makes centralized and overall cache optimization for user requests. We call the control plane of human-like hybrid caching architecture the learning module. The data plane is called the data collection module. Next, we describe the human-like hybrid caching architecture in software-defined edge cloud in detail.

\subsubsection{The flow of human-like hybrid caching}

In this paper, we propose the human-like hybrid caching architecture is shown in Fig.~\ref{fig02a}. There are two modules in the human-like caching architecture, i.e., data collection module and human-like learning module. The data collection module on the edge cloud is responsible for receiving and storing user requests in a period of time. The human-like learning module is responsible for analyzing the requested data. In the human-like learning module, classification is conducted to the content requests, and the similar requests are classified into one category. Specifically, in this paper, we use the popularity distribution of file requests to classify the content. And, the requested content will be allocated to the IRM if the popularity distribution is static. If the popularity distribution is time-varying and dynamic, then these requests are allocated to the SNM.

Furthermore, the flow of human-like hybrid caching is as follows:
we first predict the distribution and classification of the user content request, i.e., when a user requests content, an analysis of historical data is used to obtain the classification of user requests.
Then, based on the content fine-grained characteristic indicator and content popularity of IRM and SNM, we formulate an optimization model to select the file with the highest cache hit ratio, and cache it on the edge cloud. Thus, when a user requests content in next caching cycle time slot, the category of this user request is analyzed in the human-like learning module, and the corresponding contents are provided to the user. Finally, this user request will be stored in the edge cloud. Then, the above cache selection steps will be repeated.

Therefore, the human-like hybrid caching architecture is not only an integrated proactive and passive caching strategy, but also comprehensively takes into account the fine-grained characteristics of the content requested by the user. The optimal contents are cached in the edge cloud, and if a user in the communication range of the edge cloud has requests for the cached content, then the content is provided to the user faster and the user experience is improved.

\subsubsection{Human-like learning model}

From what has been discussed above, the key technology of human-like hybrid caching architecture is the learning module in control plane. The human-like learning module is shown in the Fig.~\ref{fig02b} which includes learning content characteristic and file popularity. The first is learning fine-grained characteristic of the content, then learning file popularity (i.e., IRM content or SNM content). Next, we give a detailed introduction.

First, we consider the impact of content characteristics on edge cloud caching strategy. The content characteristics have an import influence on the content requested by the user. For example, there are two content in the content library, i.e., content A and content B, and the the probability of request for contents A is higher than that of request for contents B, but the size of contents A is larger than that contents B. In addition, the transmission bandwidth of contents A is larger than that of contents B. Therefore, if we cache the content B in edge cloud, there will be more storage capacity and transmission bandwidth to cache more contents that would be requested by a user, thus improving the overall cache hit ratio.

Then, we explain how to learn the file popularity. Specifically, we need to determine whether these content requests follow a model of IRM or that of SNM, and the proportion of each model. Since the content requests in compliance with IRM and SNM may be not the same, we can not just divide the storage space of the edge cloud equally. Therefore, based on the historical data collected from the edge cloud, we use learning method to predict the proportion of IRM and SNM types, and then to predict the probability of popularity of each request. Thus, a corresponding proportion of caching space is reserved on the caching library of the edge cloud.

\section{System Model} \label{sec:System Model}

In this section, based on the content popularity model and fine-grained user characteristic model, we formulate the optimization problem aiming at the maximum cache hit ratio.

\subsection{System Overview}

In this paper, we consider a software-defined edge cloud network, which includes an edge cloud and $U$ users. The user set is defined as $\mathbb{U} = \{1,2,\cdots, U\}$. We denote $C$ as the cache capacity of the edge cloud. We consider that the file library contains $F$ files. Let $\mathbb{F} = \{1,2,\cdots, F\}$ denote the set of all files in the file library, and the size of file $f$ is $c_{f}$.  Each user will randomly and independently request files from the file library.

For the content caching, considering the limited storage capacity of edge cloud, only part of the contents can be selected to cache. If the content requested by a user is cached in edge cloud, then this user is called as a caching hit user. In this paper, our goal is to maximize cache hit ratio, that is, the proportion of caching-hit contents in the total amount of contents contentss requested by a user. If a user is not a caching-hit user, then this user will obtain contentss from the remote cloud.

Furthermore, according to the human-like hybrid caching architecture, we divide the request pattern as SNM and IRM. Let $N_S$ and $N_I$ denote the amounts of these two kinds of requests, respectively. Thus, we can separate the file library into two parts, i.e., SNM file library and IRM library. Let $\mathbb{F}^{S} = \{S_1, S_2,\cdots, S_{N_S}\}$ denote the set of all files whose request model obeys SNM, and $\mathbb{F}^I = \{I_1, I_2, \cdots, I_{N_I}\}$ denote the set of all files whose request model obeys IRM. In addition, the proportions of these two kinds of requests are recorded as $W_S$ and $W_I$, which as follows:
\begin{eqnarray}
&& W_S = \frac{\mathbb{E}[N_S]}{\mathbb{E}[N_S]+\mathbb{E}[N_I]}\\
&& W_I = \frac{\mathbb{E}[N_I]}{\mathbb{E}[N_S]+\mathbb{E}[N_I]}
\end{eqnarray}
where $\mathbb{E}[N_S], \mathbb{E}[N_I]$ is the expectation of $N_{S}$  and $N_{I}$, respectively. The reason for the expectation here is that the cache strategy estimates the number of requests of files every other period of time. Thus, when the user requests enter the human-like hybrid caching module, we first distinguish which kind of the user request belong to.

\subsection{Fine-grained Content Characteristic Model}

The fine-grained characteristic model of user requested content is as follows. When a user content request is sent to the data collection module on the edge cloud, the request will be transmitted to the human-like learning module for analysis. Then, the classification is conducted according to the user requests. Specifically, these contents can be divided into SNM contents and IRM contents, and fine-grained characteristic extraction is carried out. Thus, we can obtain the fine-grained Characteristic of contents.

The fine-grained characteristic indicators of content include the size of the user requested content, category (e.g., film, TV plays, music pieces), content transmission bandwidth, the value of the content.
Here, the value means the evaluation of requested content by users. For example, for the score of a film, and if a film is a popular, a low score of the film. Specifically, let $\Theta_{f}$ denote the characteristic indicator space of requested content $f$ with a dimension of $D$. For the fine-grained characteristic indicator of the requested content, we regularize them to the $[0, 1]$ range, as shown in the below:
\begin{equation}
\theta_{f} = \left\{ x_{f}^{i} | x_{f}^{i} \in [0,1], i=1,2,\cdots,D \right\}
\end{equation}
Thus, we use $\theta_{f}$ to represent the fine-grained characteristics of content on the human-like hybrid caching model.

\subsection{Content Popularity Model}

Based on the different request model of users for the file library, we give the cache hit ratio of SNM and IRM content. We first give the popularity and cache hit ratio of IRM content. According to the human-like hybrid caching architecture, the IRM content request pattern conform to Zipf distribution, which as follow
\begin{eqnarray}
&& \mu_{I_f} = \frac{I_f^{-\delta} }{ \sum_{j = 1}^{N_I}j^{-\delta}}
\end{eqnarray}
where the parameter $\delta$ is the popularity skewness.

Furthermore, we denote $y_{f}^{I}$ as whether the IRM content ${I_f}$ requested by the user is to be cached in the edge cloud. If $y_{f}^{I}=1$, the IRM content $I_{f}$ is cached in the edge cloud. If $y_{f}^{I}=0$, the IRM content is not cached in the edge cloud. According to~\cite{14}, the cache hit ratio of IRM content can be obtained as follows:
\begin{equation}\label{eq:3}
\begin{split}
&P_{hit}^I = \sum_{f=1}^{N_I} y_{f}^{I} \times \mu_{I_f}(\theta_{I_f})
\end{split}
\end{equation}
Thus, we obtain the cache hit ratio of IRM content. From the~\eqref{eq:3}, we can get that the more popularity of IRM files, the higher the cache hit ratio of content.

Then, for the SNM content, let $V$ denote the number of requests for SNM content $S_{f}$, which is assumed to be independent identically distributed. According to~\cite{13}, $V$ conforms to the following Pareto distribution:
\begin{eqnarray}
&& f (v) = \beta N_{min}^\beta v^{-(\beta+1)}
\end{eqnarray}
where $N_{min}$ is the minimum number of requests corresponding to all SNM requested contents, $v\geq N_{min}$ and parameters $\beta(\beta>1)$ is the popularity skewness. Furthermore, in order to calculate the cache hit ratio for contents requested by a user that conform to SNM, we define $\lambda(t)$ to express the content request rate over time, and the expectation of $V$, $\mathbb{E}[V]<\infty$. In addition, we define $\phi_V(x)=\mathbb{E}[e^{xV}]$ to represent the moment generating function of $V$, and the first-order derivative is shown as follows:
\begin{equation}\label{eq:4}
\phi'_V(x) = \mathbb{E}[Ve^{xV}], \quad \phi_V(x) \geq 0
\end{equation}
Reference to Che's approximation~\cite{15}, the traditional computing method of cache hit ratio for the SNM is as follows:
\begin{equation}\label{eq:5}
\begin{split}
& P_{hit}^S = 1- \int_{0}^{\infty}\lambda(\tau) \frac {\phi_{V_f}' \left( -\int_{0}^{T_C} \lambda (\tau - \eta) d \eta \right)}{\mathbb{E}[V]} d \tau
\end{split}
\end{equation}
where, $T_C$ is the unique solution to the following equation:
\begin{equation}\label{eq:6}
\begin{split}
& C = \gamma \int_{0}^{\infty} 1-\phi_{V_f} \left( -\int_{0}^{T_C} \lambda (\tau - \eta) d \eta \right)
\end{split}
\end{equation}
However, this calculation method is complex and does not take into account fine-grained content characteristics, Thus, in this paper, we use online algorithms to learn the popularity of SNM.

Furthermore, we define $y_{f}^{S}$ to indicate whether the SNM content $S_{f}$ requested by the user is to be cached in the edge cloud. If $y_f^S=1$, the content $S_{f}$ is cached in the edge cloud. If $y_{f}^{S}=0$, the content is not cached in the edge cloud. Then we can obtain cache hit ratio of SNM contents as follows:
\begin{equation}\label{eq:7}
\begin{split}
P_{hit}^S = \sum_{f = 1}^{N_S} y_f^S \times \mu_{S_f}(\theta_{S_f})
\end{split}
\end{equation}
where $\mu_{S_f}(\theta_{S_f})$ is the popularity of the SNM content $S_{f}$, which can be obtained through online algorithm.

\subsection{Problem Formulation}

We formulate the human-like hybrid content caching problem. First, based on the cache hit ratio of IRM content and SNM content, we can obtain the cache hit ratio of all files as follows:
\begin{equation}\label{eq:8}
\begin{split}
P_{hit} &= \sum_{f=1}^{F}\left( W_I P_{hit}^I + W_S P_{hit}^S \right) \\
& =  \sum_{f\in \mathbb{F}^I} W_I y_f^I  \mu_{I_f} (\theta_{I_f}) + \sum_{f\in \mathbb{F}^S} W_S y_f^S  \mu_{S_f}(\theta_{S_f})
\end{split}
\end{equation}

Furthermore, the human-like hybrid caching optimization problem can be obtained as follows:
\begin{eqnarray}
\label{eq:10}
\bf{P1:}  \underset{y_f^I,y_f^S}{\text{maximize}}
&&   P_{hit} \\
\text{subject to}
&& C1: y_f^I,y_f^S \in \{0,1\}.\\
&& C2: \sum_{f\in \mathbb{F}^I} c_f y_f^I + \sum_{f\in \mathbb{F}^S} c_f y_f^S \leq C.
\end{eqnarray}
where the objective function is to maximize the cache hit ratio. The first constraint C1 indicates that the $y_{f}^{I}$ and $y_{f}^{S}$ are 0 or 1. The C2 constraint indicates that the cached content can not exceed the total storage of edge cloud.

For the optimization problem $\mathbf{P1}$, if we already know the proportion of IRM content and SNM content, and the corresponding popularity, the optimization problem $\mathbf{P1}$ is a typical 0-1 integer optimization which can be solved. However, in practice, the popularity of SNM contents is time-varying, and the cache hit ratio of SNM content is related to the fine-grained characteristic indexes of user requested, which need online learning. Thus, in order to solve the optimization problem $\mathbf{P1}$, we use the multi-arm bandit theory give the human-like hybrid caching scheme.

\section{Human-like Hybrid Caching Scheme} \label{sec:solution}

In this section, we use a combination of neural networks and reinforcement learning to solve optimization problem $\mathbf{P1}$, and propose the human-like hybrid caching scheme.

\begin{algorithm}[!htb]
\caption{Human-like hybrid caching algorithm}
\label{alg:Framworka}
\begin{algorithmic}[1]
\REQUIRE ~~ $T$,  $\mathbb{F}$\\

    \STATE Initialization: for all $f\in {\mathbb{F}}$,  set $N_{f}=0$ , $P_{hit}^S(t)=0$,
    $r_f$, $A = 0$, $B_{r_f}=\mathrm{Bern}(r_f)$, where $\mathrm{Bern}$ is Bernoulli distribution, and the fine-grained characteristic is $\theta$.

\FOR{$t= 1, \cdots, T$}

     \STATE  Observe the candidate subset of file library $\mathbb{F}^S(t)$ at time slot $t$, and the number of $\mathbb{F}^S(t)$ is $N_{t}$, i.e., $\vert{\mathbb{F}^S(t)}\vert = N_t$
     \IF{ any content $f\in{\mathbb{F}^S(t)}$ has not been cached}
         \STATE Caching content $f$ once.
         \STATE Update the caching hit ratio $P_{hit}^S(t)$, and the $x_{f,t}^S$ according to the observation
         \STATE Update $r_{t,f}=1- P_{hit}^S(t,f)/ P_{hit,max}^S(t,f)$
         \STATE Update $A_{t,f}=1$
         \STATE Update $B_{r_{t,f}}=A_{t,f}\times r_{t,f}$
         \STATE Update $N_{f}(t)=N_{f}(t)+1$
     \ELSE
         \STATE Observe fine-grained characteristic indexes $x_{f,t}$ and caching hit ratio $P_{hit}^S(t)$
         \STATE Calculate the caching hit ratio according to
         \begin{equation}\label{eq:15}
           \tilde P_{hit}^S(t,f) =
            \mathrm{\bar{P}}_{hit}^S(t-1,f)+\sqrt{\tfrac{\beta W_{r_{t-1,f}}x_{f,S}^t\ln(t)}{N_{f}(t-1)}}
         \end{equation}

         \WHILE{$\sum_{f\in \mathbb{F}^I} c_f y_{f,t}^I + \sum_{f\in \mathbb{F}^S} c_f y_{f,t}^S \leq C$}
             \STATE Select content $f$ to cache according to
             \begin{equation}\label{eq:16}
             \begin{split}
              y_{f,t}^S &=\mathop{\arg \max}_{f\in \mathbb{F}^S(t)} \left(W_S \tilde P_{hit}^S(t,f)+W_I P_{hit}^I\right) \\
             \end{split}
             \end{equation}
             \STATE Update $r_{t,f} = P_{hit}^S(t,f)/ P_{hit,max}^S(t,f)$
             \STATE Update $A_{t,f}=1$ and the $x_{f,t}^S$ according to the observation
              \STATE Update $B_{r_{t,f}}=A_{t,f}r_{t,f}$
             \STATE Update $N_{f}(t)=N_{f}(t-1)+1$
             \STATE Update $\mathrm{\bar{P}}_{hit}^S(t,f)=\frac{\mathrm{\bar{P}}_{hit}^S(t-1,f) N_{f}(t-1)+P_{hit}^S(t,f)}{N_{f}(t)}$
         \ENDWHILE
     \ENDIF
\ENDFOR
\end{algorithmic}
\end{algorithm}

\subsection{Cache Capacity Allocation Prediction}

For the cache capacity allocation prediction, we adopt a neural network model to predict which model a user request conforms to. Specifically, we first divide $T_C$ into $T$ time periods, and all the user requested contents in the caching cycle are collected. The user content requests are classified according to the distribution types of the IRM and the SNM, and $N_S^t$ and $N_I^t$ are calculated. Then, we can obtain $W_S^t$ and $W_I^t$. Finally, these historical data are made as training datasets. The trained neural network model is adopted to predict the probability of IRM contetn, and the predicted probability is made according to the estimated value of $W_I$ and $W_S = 1-W_I$. In this paper, we use the neural network model proposed in~\cite{15} to predict the proportion of the cache capacity allocation. Thus, we can obtain the $W_{I}$ and $W_{S}$.

\subsection{Human-like Hybrid Caching Algorithm}

We present the analysis of the caching scheme from the IRM content and the SNM content. First, for the IRM content, due to the long life cycle of the IRM file library, the file popularity changes little over time, so for IRM content, we select the cache content directly by popularity sort.
As for the SNM contents, the life cycle is short, and the total request number is large. Thus, we can divide the cache cycle period  into small time periods, where the number change in SNM content can be ignored. Therefore, we use reinforcement learning to select cached files. Specifically, in this paper, we design a human-like hybrid caching algorithm by learning fine-grained characteristic and content popularity, where the influence of the fine-grained characteristic and content popularity on content selection are taken into consideration. Next, we describe the human-like hybrid caching algorithm in detail.

As shown in the Algorithm 1, first, we make $T$ and $\mathbb{F}$ as the input of the algorithm, where $T$ represents the number of time periods divided by the entire cache cycle. In the initialization stage of the algorithm, we define the $N_f$ as the number of times of caching the content $f$. The weight of cache hit ratio of content $f$ is defined as $r_f$, which is a random number between [0,1]. $r_{t,f}$ is the value of $r_f$ at time slot $t$. In addition, we set up a fine-grained characteristic $\theta$, and the element at time slot $t$ is $x_{f, t}^S$. This indicates the degree to which the fine-grained characteristic indexes of the requested content of $f$ at the time slot $t$ have an impact on the cache hit ratio. Furthermore, we set the cache hit ratio $P^S_{hit, f} = 0 $. The parameters $A$ and $B_{r_f} $ are the coefficients of the effect of requested content on cache hit ratio. In fact, $A $ is equivalent to a actions in reinforcement learning. $A_{t,f} $ and $B_ {r_{t,f}} $ are the value of $A$ and $B_ {r_f}$ at time slot $t$, respectively.

Second, we need to get the initial cache hit ratio for each candidate content in the file library. So we record the cache hit ratio of SNM content in the SNM file library $\mathbb{F}^S(t)$ at the first time slot, and update $x_{f,t}^S$, $\mathrm{\bar{P}}_{hit}^S(t,f)=P_{hit}^S(t,f) $, where $\mathrm{\bar{P}}_{hit}^S(t,f)$ is the average cache hit ratio of the selected cache content. Then update the parameter as follows:
\begin{equation}\label{eq:17}
\begin{split}
&r_{t,f}= P_{hit}^S(t,f)/ P_{hit,max}^S(t,f) \\
&A_{t,f}=1, B_{r_{t,f}}=A_{t,f}\times r_{t,f}\\
&N_{f}(t)=N_{f}(t)+1
\end{split}
\end{equation}

After the first time slot, the cache content $f$ is selected according to~\eqref{eq:15} and~\eqref{eq:16}, where $\mathrm{\bar{P}}_{hit}^S(t,f)+\sqrt{\tfrac{\beta W_{r_{t,f}} x_{f,S} ^ t \ln(t)} {N_{f}(t-1)}}$ is used to balance the exploration and exploitation of this algorithm. From the~\eqref{eq:16}, it can be concluded that when the cache hit ratio $\mathrm{\bar{P}}_{hit}^S(t,f)$ of file $f$ is high or the exploration times $N_{f}(t-1)$ are low, the cache probability of this file is higher.  Since the IRM content popularity changes slowly, we select $y_{f}^{I}$ based on the popularity of the IRM content. Thus, the~\eqref{eq:16} is equivalent to the following
\begin{equation}\label{eq:18}
y_{f}^{S}=\mathop{\arg \max}_{f\in \mathbb{F}^S(t)} \tilde P_{hit}^S(t,f)
\end{equation}

Finally, we use~\eqref{eq:17} and~\eqref{eq:19} to update every parameter, as shown below:
\begin{equation}\label{eq:19}
\begin{split}
\mathrm{\bar{P}}_{hit}^S(t,f)
&=\frac{\mathrm{\bar{P}}_{hit}^S(t-1,f) N_{f}(t-1)+P_{hit}^S(t,f)}{N_{f}(t)}.
\end{split}
\end{equation}
Furthermore, when the file size of the cache exceeds the edge cloud storage capacity, the cache placement is terminated.

\section{Performance Evaluation}
\label{sec.performance}

In this section, we present the simulation experiment and verify the human-like hybrid algorithm.

\subsection{Simulation Setup and Comparison Algorithm }

First, we give the system simulation settings. Specifically, we assume that the total caching cycle is divided into 600 time slots, i.e, $T=600$. For the content library, we set the number of contents in the file library to $|\mathbb{F}|=150$. The total number of contents belong to SNM is $N_S = W_S|\mathbb{F}|$. Since the total number for requests for SNM content is larger than IRM content, according to the cache allocation model, we obtain $W_S=0.8$. Furthermore, we set the cache capacity of edge cloud $C = 30$. For the Algorithm 1, we set the parameters $\beta=2$.

Furthermore, for the content popularity, since the file library of the SNM is time-varying, a single content request data could use the Pareto distribution to simulate the cache hit rate of 600 time periods. Furthermore, according to~\cite{15}, the popularity of a single content per time period is consistent with the Zipf distribution. For content fine-grained characteristics, we use real data sets~\cite{16}. Thus, we can obtain the content popularity of all requests for a time slot.

Then, we compare human-like hybrid caching algorithm with random caching and popular caching algorithms. The details are as follows:
\begin{itemize}
\item \emph{Random caching algorithm}. The random caching algorithm randomly select content to cache in edge cloud. Thus, this algorithm takes neither the fine-grained characteristic of user nor the content popularity into account.

\item \emph{Popular caching algorithm~\cite{5}}. The popular caching algorithm select the most popularity content to cache in the edge  cloud. However, this algorithm does not consider the fine-grained characteristics of user.

\end{itemize}

\subsection{Impact of File Library}

\begin{figure*}[!ht]
\centering
\subfigure[Cache hit ratio under different size of file library.]
{\includegraphics[width=3.5in]{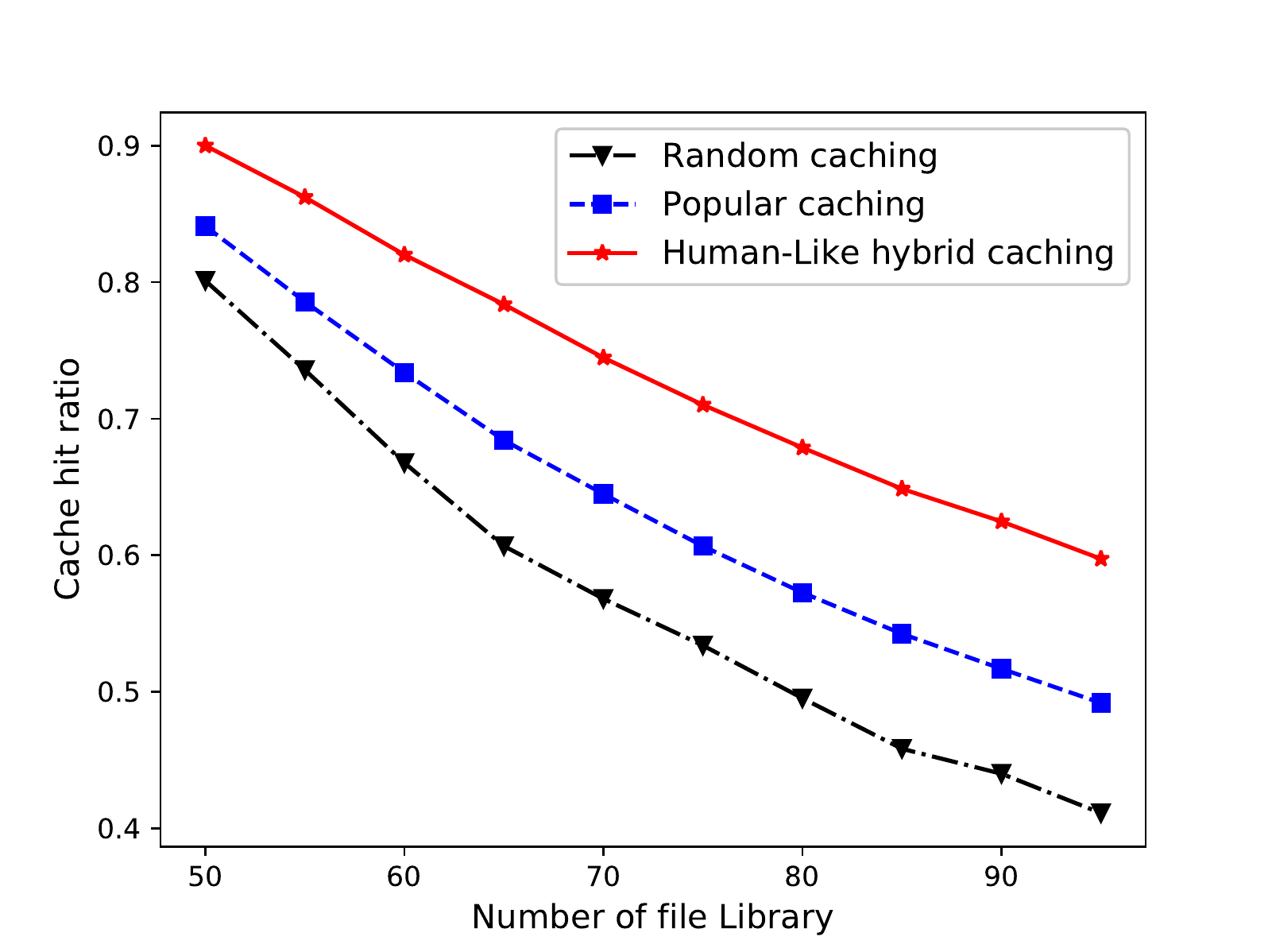}
\label{fig005a}
}
\subfigure[Cumulative learning regret under different size of file library.]
{\includegraphics[width=3.5in]{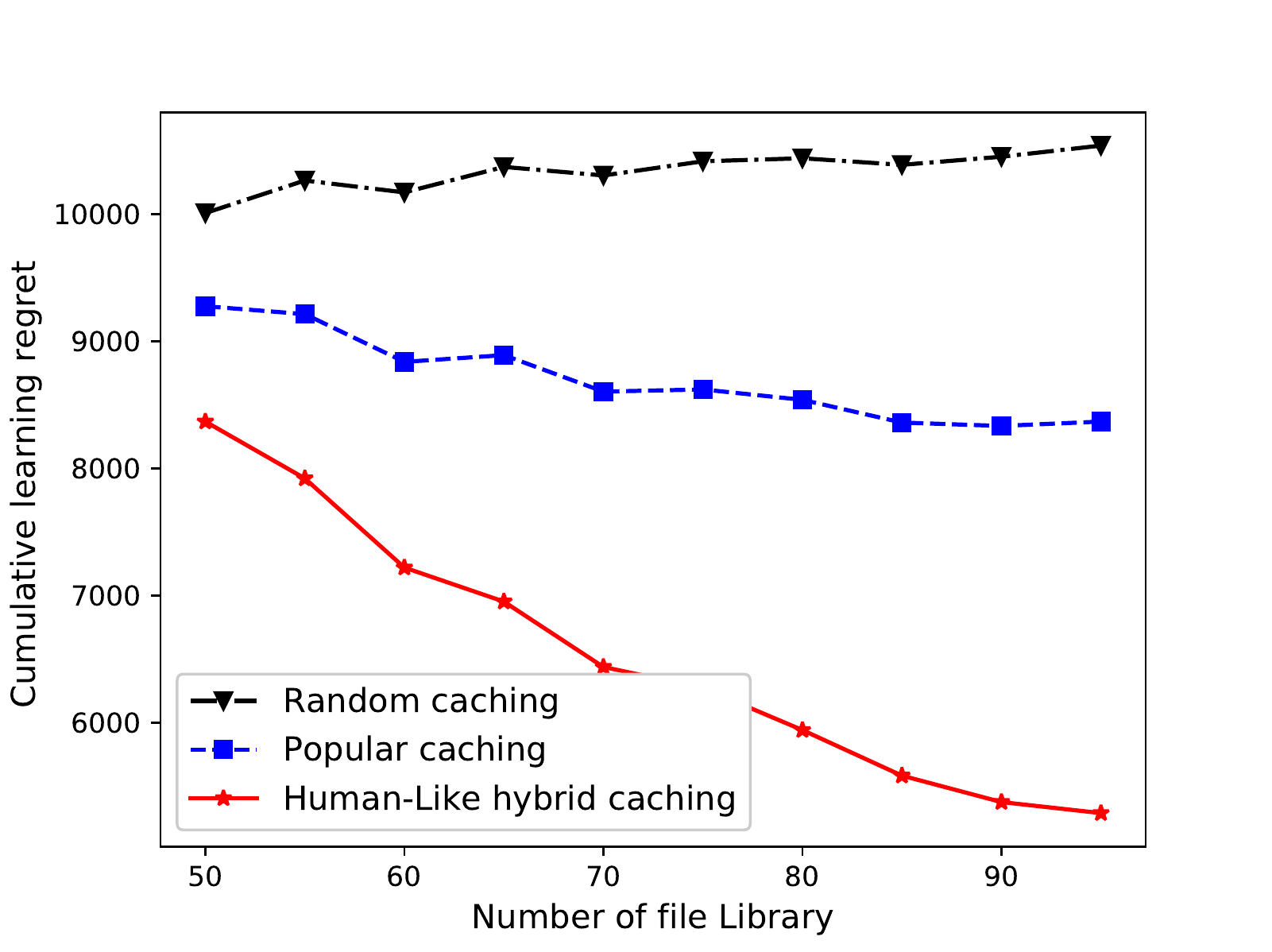}
\label{fig005b}
}
\caption{The impact of file library on cumulative learning regret and cache hit ratio.}
\label{fig005}
\end{figure*}

First, we evaluated the cache hit ratio and cumulative learning regret under different number of file library, as shown in Fig.~\ref{fig005}. The cumulative regret refers to the difference between the benefit of each selected cache content and the benefit of the file with the highest cache hit ratio~\cite{chen2016combinatorial}. In this experiment, we set $T=600$, and the cache capacity is 40. Fig.~\ref{fig005} indicates that the human-like hybrid caching algorithm is significantly better than the random and popular cache algorithms in cache hit ratio and cumulative learning regret.
For example, in terms of cache hit rate, when the size of the file library is 90, compared with random cache algorithm and popular algorithm, the human like cache strategy is improved by 40\% and 20\% respectively. This is because human-like hybrid caching algorithm not only considers the content popularity, but also considers the fine-grained characteristic of users.

Furthermore, as shown in Fig.~\ref{fig005a}, with the increase of the number of file library, the cache hit ratio of the human-like hybrid cache algorithm decreases. This is because with the increase of the file library, the number of files that users can request increases, but the cache capacity of the edge cloud does not change. Moreover, Fig.~\ref{fig005b} indicates that with the increase of the number of file libraries, the cumulative learning regret of human-like hybrid caching decreases gradually. Thus, through Fig.~\ref{fig005}, we can obtain that the human-like hybrid caching algorithm is more practical because the file library in reality is generally large.

\subsection{Impact of Caching Capacity}

\begin{figure*}[!ht]
\centering
\subfigure[Cache hit ratio under different caching capacity.]
{\includegraphics[width=3.5in]{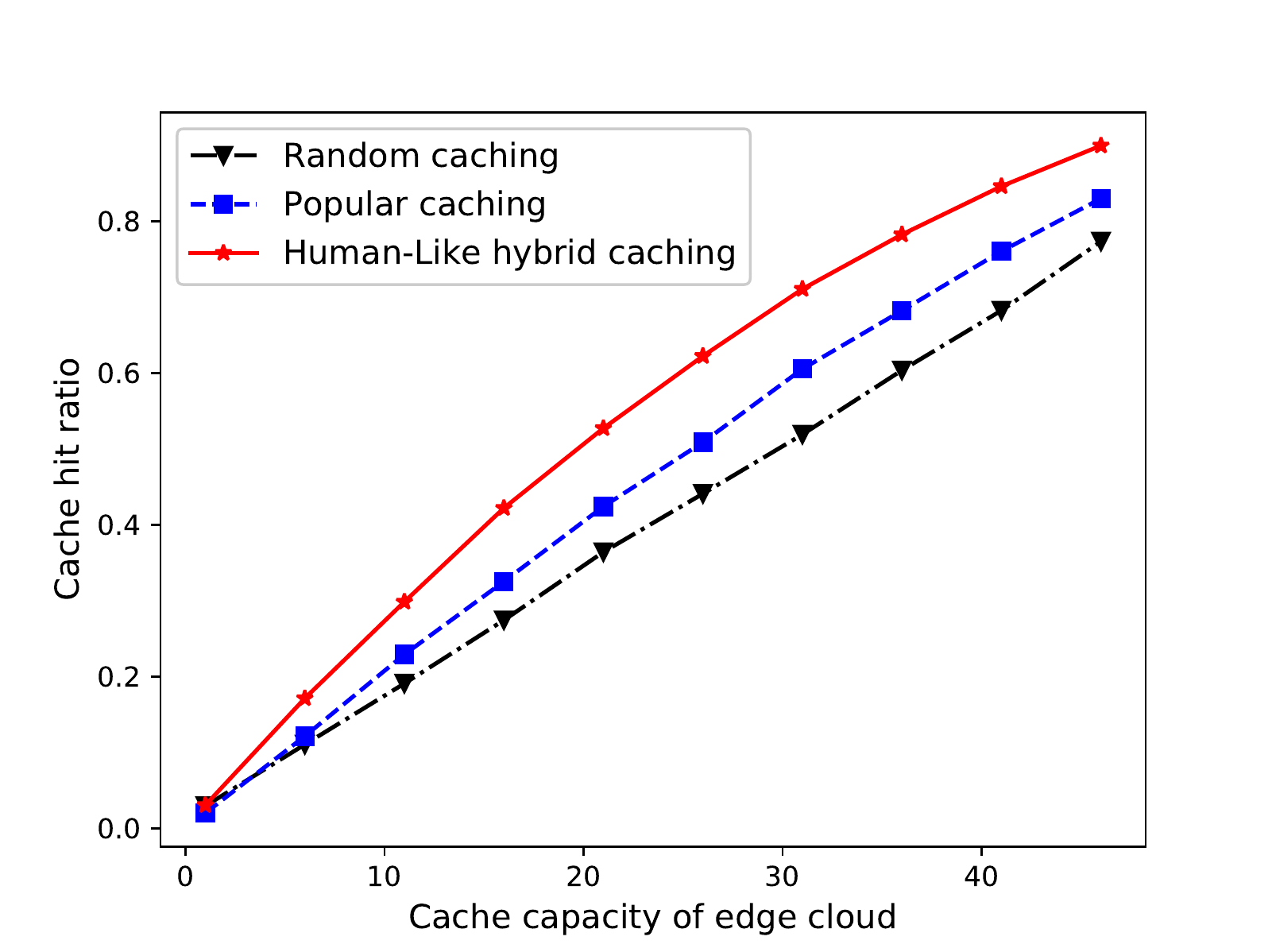}
\label{fig006a}
}
\subfigure[Cumulative learning regret under different caching capacity.]
{\includegraphics[width=3.5in]{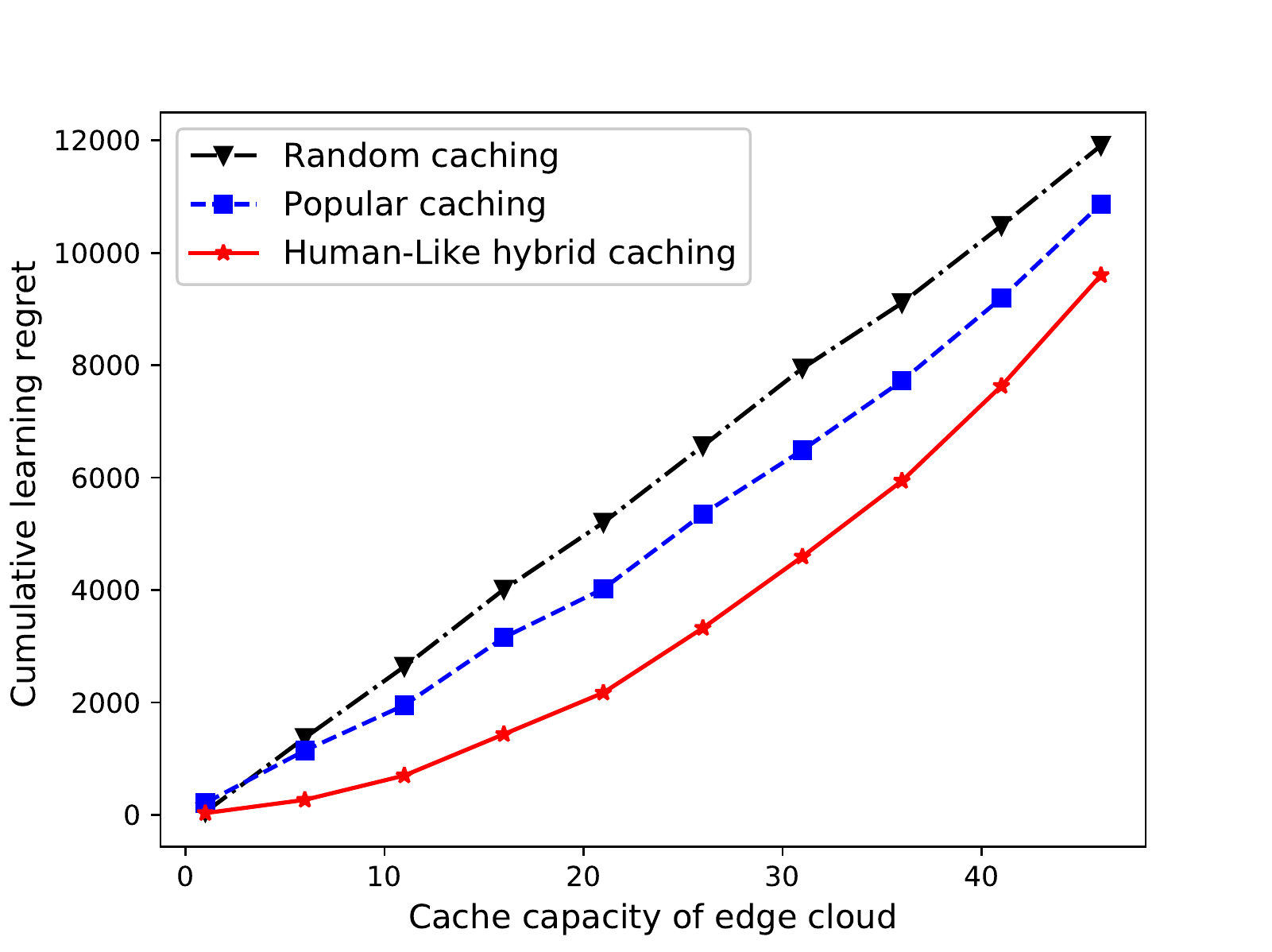}
\label{fig006b}
}
\caption{The impact of caching capacity on cumulative learning regret and cache hit ratio.}
\label{fig006}
\end{figure*}

Then, we study the impact of cache capacity on the cache hit ratio and cumulative learning regret, as shown in the Fig.~\ref{fig006}. In this experiment, we set the size of the file library to be 150, and obtain the cache hit ratio and cumulative learning regret corresponding to different cache capacity at $T=600$. From Fig.~\ref{fig006}, we can see that the human-like hybrid caching algorithm has higher cache hit ratio and lower regret under different cache capacity.

Furthermore, from Fig.~\ref{fig006a}, we can see that the cache hit ratio became higher as the cache capacity became bigger. This is because as the size of cache capacity increases, the edge cloud can cache more content, thus improving the cache hit ratio. Moreover, as shown in Fig.~\ref{fig006b}, the cumulative learning regret increases with the cache capacity of edge cloud. This can be explained as more cache files are added with the increase of cache capacity, and every cache file will bring learning regret. Therefore, the human-like hybrid algorithm is superior to random and popular algorithms under different size of cache capacity.

\section{Conclusion}\label{sec.conclusion}

In this paper, we first introduce the human-like hybrid caching  architecture in software-defined edge cloud. Then, in view of the content popularity model and user fine-grained characteristic model, we formulate the optimization problem of maximum cache hit ratio. In order to solve this problem, we propose a hybrid human-like algorithm based on reinforcement learning. Extensive experiments show that the human-like caching algorithm significantly outperforms the random caching algorithm and popular caching algorithm. Furthermore, we show the impact of file library and cache capacity on cache hit ratio. The result shows that our scheme is 20\% better than the popular caching algorithm. In future work, we will consider both user and content fine-grained to further improve the cache hit ratio.

\section*{Acknowledgement}

The authors are grateful to the Deanship of Scientific Research at King Saud University for funding this work through the Vice Deanship of Scientific Research Chairs: Chair of Smart Technologies.

\bibliographystyle{IEEEtran}

\bibliography{mybibfile}

\begin{thebibliography}{10}
\providecommand{\url}[1]{#1}
\csname url@samestyle\endcsname
\providecommand{\newblock}{\relax}
\providecommand{\bibinfo}[2]{#2}
\providecommand{\BIBentrySTDinterwordspacing}{\spaceskip=0pt\relax}
\providecommand{\BIBentryALTinterwordstretchfactor}{4}
\providecommand{\BIBentryALTinterwordspacing}{\spaceskip=\fontdimen2\font plus
\BIBentryALTinterwordstretchfactor\fontdimen3\font minus
  \fontdimen4\font\relax}
\providecommand{\BIBforeignlanguage}[2]{{%
\expandafter\ifx\csname l@#1\endcsname\relax
\typeout{** WARNING: IEEEtran.bst: No hyphenation pattern has been}%
\typeout{** loaded for the language `#1'. Using the pattern for}%
\typeout{** the default language instead.}%
\else
\language=\csname l@#1\endcsname
\fi
#2}}
\providecommand{\BIBdecl}{\relax}
\BIBdecl

\bibitem{1}
S.~Garg, K.~Kaur, G.~Kaddoum, S.~Ahmed, and D.~N. Jayakody, ``Sdn based secure
  and privacy-preserving scheme for vehicular networks: A 5g perspective,''
  \emph{IEEE Transactions on Vehicular Technology}, vol.~PP, pp. 1--1, 05 2019.

\bibitem{fortino2008using}
G.~Fortino and W.~Russo, ``Using p2p, grid and agent technologies for the
  development of content distribution networks,'' \emph{Future Generation
  Computer Systems}, vol.~24, no.~3, pp. 180--190, 2008.

\bibitem{sun2019tide}
P.~Sun, Y.~Hu, J.~Lan, L.~Tian, and M.~Chen, ``Tide: Time-relevant deep
  reinforcement learning for routing optimization,'' \emph{Future Generation
  Computer Systems}, vol.~99, pp. 401--409, 2019.

\bibitem{chen2016opportunistic}
M.~Chen, Y.~Hao, C.-F. Lai, D.~Wu, Y.~Li, and K.~Hwang, ``Opportunistic task
  scheduling over co-located clouds in mobile environment,'' \emph{IEEE
  Transactions on Services Computing}, vol.~11, no.~3, pp. 549--561, 2016.

\bibitem{fortino2014agent}
G.~Fortino, W.~Russo, and M.~Vaccaro, ``An agent-based approach for the design
  and analysis of content delivery networks,'' \emph{Journal of Network and
  Computer Applications}, vol.~37, pp. 127--145, 2014.

\bibitem{4}
X.~Tan, Y.~Guo, Y.~Chen, and W.~Zhu, ``Characterizing user popularity
  preference in a large-scale online video streaming system,'' 2015.

\bibitem{5}
N.~{Golrezaei}, K.~{Shanmugam}, A.~G. {Dimakis}, A.~F. {Molisch}, and
  G.~{Caire}, ``Femtocaching: Wireless video content delivery through
  distributed caching helpers,'' in \emph{2012 Proceedings IEEE INFOCOM}, March
  2012, pp. 1107--1115.

\bibitem{6}
Y.~Guo, L.~Duan, and R.~Zhang, ``Cooperative local caching under heterogeneous
  file preferences,'' \emph{IEEE Transactions on Communications}, vol.~65,
  no.~1, pp. 444--457, 2016.

\bibitem{2}
M.~Hefeeda and O.~Saleh, ``Traffic modeling and proportional partial caching
  for peer-to-peer systems,'' \emph{IEEE/ACM Transactions on networking},
  vol.~16, no.~6, pp. 1447--1460, 2008.

\bibitem{3}
S.~Traverso, M.~Ahmed, M.~Garetto, P.~Giaccone, E.~Leonardi, and S.~Niccolini,
  ``Temporal locality in today's content caching: why it matters and how to
  model it,'' \emph{ACM SIGCOMM Computer Communication Review}, vol.~43, no.~5,
  pp. 5--12, 2013.

\bibitem{15}
K.~Qi, S.~Han, and C.~Yang, ``Learning a hybrid proactive and reactive caching
  policy in wireless edge under dynamic popularity,'' \emph{IEEE Access},
  vol.~7, pp. 120\,788--120\,801, 2019.

\bibitem{7}
C.~Koch, S.~Werner, A.~Rizk, and R.~Steinmetz, ``Mira: Proactive music video
  caching using convnet-based classification and multivariate popularity
  prediction,'' in \emph{2018 IEEE 26th International Symposium on Modeling,
  Analysis, and Simulation of Computer and Telecommunication Systems
  (MASCOTS)}.\hskip 1em plus 0.5em minus 0.4em\relax IEEE, 2018, pp. 109--115.

\bibitem{8}
Y.~Jiang, M.~Ma, M.~Bennis, F.-C. Zheng, and X.~You, ``User preference
  learning-based edge caching for fog radio access network,'' \emph{IEEE
  Transactions on Communications}, vol.~67, no.~2, pp. 1268--1283, 2018.

\bibitem{9}
B.~Bharath, K.~G. Nagananda, and H.~V. Poor, ``A learning-based approach to
  caching in heterogenous small cell networks,'' \emph{IEEE Transactions on
  Communications}, vol.~64, no.~4, pp. 1674--1686, 2016.

\bibitem{12}
H.~Farahat, R.~Atawia, and H.~S. Hassanein, ``Robust proactive mobility
  management in named data networking under erroneous content prediction,'' in
  \emph{GLOBECOM 2017-2017 IEEE Global Communications Conference}.\hskip 1em
  plus 0.5em minus 0.4em\relax IEEE, 2017, pp. 1--6.

\bibitem{chen2019cognitive}
M.~Chen, Y.~Hao, H.~Gharavi, and V.~C. Leung, ``Cognitive information
  measurements: A new perspective,'' \emph{Information Sciences}, vol. 505, pp.
  487--497, 2019.

\bibitem{10}
S.~M{\"u}ller, O.~Atan, M.~van~der Schaar, and A.~Klein, ``Context-aware
  proactive content caching with service differentiation in wireless
  networks,'' \emph{IEEE Transactions on Wireless Communications}, vol.~16,
  no.~2, pp. 1024--1036, 2016.

\bibitem{11}
T.~Spyropoulos and P.~Sermpezis, ``Soft cache hits and the impact of
  alternative content recommendations on mobile edge caching,'' in
  \emph{Proceedings of the eleventh ACM workshop on challenged networks}.\hskip
  1em plus 0.5em minus 0.4em\relax ACM, 2016, pp. 51--56.

\bibitem{aben2012distinction}
B.~Aben, S.~Stapert, and A.~Blokland, ``About the distinction between working
  memory and short-term memory,'' \emph{Frontiers in psychology}, vol.~3, p.
  301, 2012.

\bibitem{chen2019label}
M.~Chen and Y.~Hao, ``Label-less learning for emotion cognition,'' \emph{IEEE
  transactions on neural networks and learning systems}, 2019.

\bibitem{14}
S.-E. Elayoubi and J.~Roberts, ``Performance and cost effectiveness of caching
  in mobile access networks,'' in \emph{Proceedings of the 2nd ACM Conference
  on Information-Centric Networking}.\hskip 1em plus 0.5em minus 0.4em\relax
  ACM, 2015, pp. 79--88.

\bibitem{13}
G.~S. Paschos, G.~Iosifidis, M.~Tao, D.~Towsley, and G.~Caire, ``The role of
  caching in future communication systems and networks,'' \emph{IEEE Journal on
  Selected Areas in Communications}, vol.~36, no.~6, pp. 1111--1125, 2018.

\bibitem{16}
L.~Chen, J.~Xu, S.~Ren, and P.~Zhou, ``Spatio--temporal edge service placement:
  A bandit learning approach,'' \emph{IEEE Transactions on Wireless
  Communications}, vol.~17, no.~12, pp. 8388--8401, 2018.

\bibitem{chen2016combinatorial}
W.~Chen, W.~Hu, F.~Li, J.~Li, Y.~Liu, and P.~Lu, ``Combinatorial multi-armed
  bandit with general reward functions,'' in \emph{Advances in Neural
  Information Processing Systems}, 2016, pp. 1659--1667.

\end{thebibliography}



\end{document}